\documentclass[12pt,preprint]{aastex}

\shorttitle{\indent \def EOF Borland loops} \shortauthors{Bourouaine
et al.}

\begin{document}

\title{Correlations between the proton temperature anisotropy and transverse high-frequency waves in the solar wind}

\author{Sofiane Bourouaine\altaffilmark{1}, Eckart Marsch\altaffilmark{1} and Fritz M. Neubauer\altaffilmark{2}}
\altaffiltext{1} {Max-Planck-Institut f\"ur Sonnensystemforschung,
37191 Katlenburg-Lindau, Germany}

\altaffiltext{2}{Institut f\"{u}r Geophysik und Meteorologie, Universit\"{a}t zu K\"{o}ln, Albertus-Magnus-Platz, K\"{o}ln, 50923, Germany}




\email{bourouaine@mps.mpg.de}

\begin{abstract}
Correlations are studied between the power density of transverse
waves having frequencies between $0.01$ and $1$ normalized to the proton
gyrofrequency in the plasma frame and the ratio of the perpendicular and parallel temperature of
the protons. The wave power spectrum is evaluated from high-resolution 3D
magnetic field vector components, and the ion temperatures are derived from
the velocity distribution functions as measured in fast solar wind during the
Helios-2 primary mission at radial distances from the Sun between 0.3~AU and
0.9~AU. From our statistical analysis, we obtain a striking correlation
between the increases in the proton temperature ratio and enhancements in the
wave power spectrum. Near the Sun the transverse part of the wave power is
often found to be by more than an order of magnitude higher than its
longitudinal counterpart. Also the measured ion temperature anisotropy
appears to be limited by the theoretical threshold value for the
ion-cyclotron instability. This suggests that high-frequency
Alfv\'{e}n-cyclotron waves regulate the proton temperature anisotropy.
\end{abstract}

\section{Introduction}

Several decades ago, the Helios in-situ measurements made in fast solar wind
already showed that the proton temperature ratio, $T_{\perp}/T_{\parallel}$,
in particular that of the core part studied here, was not unity but
could reach high values of up to 3 at a distance of 0.3 AU from the Sun
\citep{marsch1982}, thus indicating strong local ion heating. With increasing
heliocentric distance, this ratio tends to decrease, attaining values equal
to unity or below only at 1~AU \citep{kasper2002}.

Many theoreticians tried to explain the observed strong perpendicular ion
heating, e.g., \cite{tu2001, marsch2001} for the Helios results,
\cite{cranmer2009} and the many authors referenced therein for the fast solar
wind. It is now widely believed that such kinetic features are in fact
signatures of ion heating via ion-cyclotron resonance. However, it has also
been questioned \citep[e.g.,][]{cranmer2001} whether ion-cyclotron waves do
really exist near the Sun with sufficient power to provide the heating, or to
enable the plateau formation in the proton velocity distribution functions
(VDFs) observed in fast solar wind \citep{tu2002}.

There is no general agreement yet on the physical process capable of
explaining the heating of the ions and their temperature anisotropy.
Recently, \cite{bale2009} showed statistically that the gyroscale
magnetic fluctuations at 1~AU are enhanced along the temperature anisotropy
thresholds of the mirror, proton oblique firehose, and ion cyclotron
instabilities which all seem to regulate the anisotropy. A possible wave
generation scenario is that a nonlinear parallel cascade (via parametric decay) of
low-frequency Alfv\'{e}n waves may ultimately generate ion-cyclotron waves,
whereby the protons and heavy ions can be heated perpendicularly via
cyclotron-resonant wave absorption \citep{araneda2008, araneda2009,
valentini2009}. In contrast to these models, it was argued that the energy
density of the Alfv\'{e}nic turbulence does not effectively cascade to the
high-frequency parallel cyclotron waves but rather goes to low-frequency
oblique kinetic Alfv\'{e}n waves (with $k_\parallel \ll k_\perp$). Then
dissipation would still take place at the proton gyroscale, with
$k_\perp=\Omega_p/V_A$, but via Landau damping acting mostly on electrons
\citep{howes2008}. Here $\Omega_p= eB/(m_pc)$ is the proton gyrofrequency,
$B$ the magnetic field strength, $e$ the elementary charge unit, $c$ the
vacuum speed of light, $m_p$ the proton mass, and $V_A= B/\sqrt{4\pi n_p
m_p}$ the Alfv\'en speed involving the proton number density $n_p$. However,
for ions Landau damping would lead to parallel heating, and hence in this way
their temperature anisotropy in fast streams can not be explained.

\begin{figure}[!U]
\includegraphics[width=40pc]{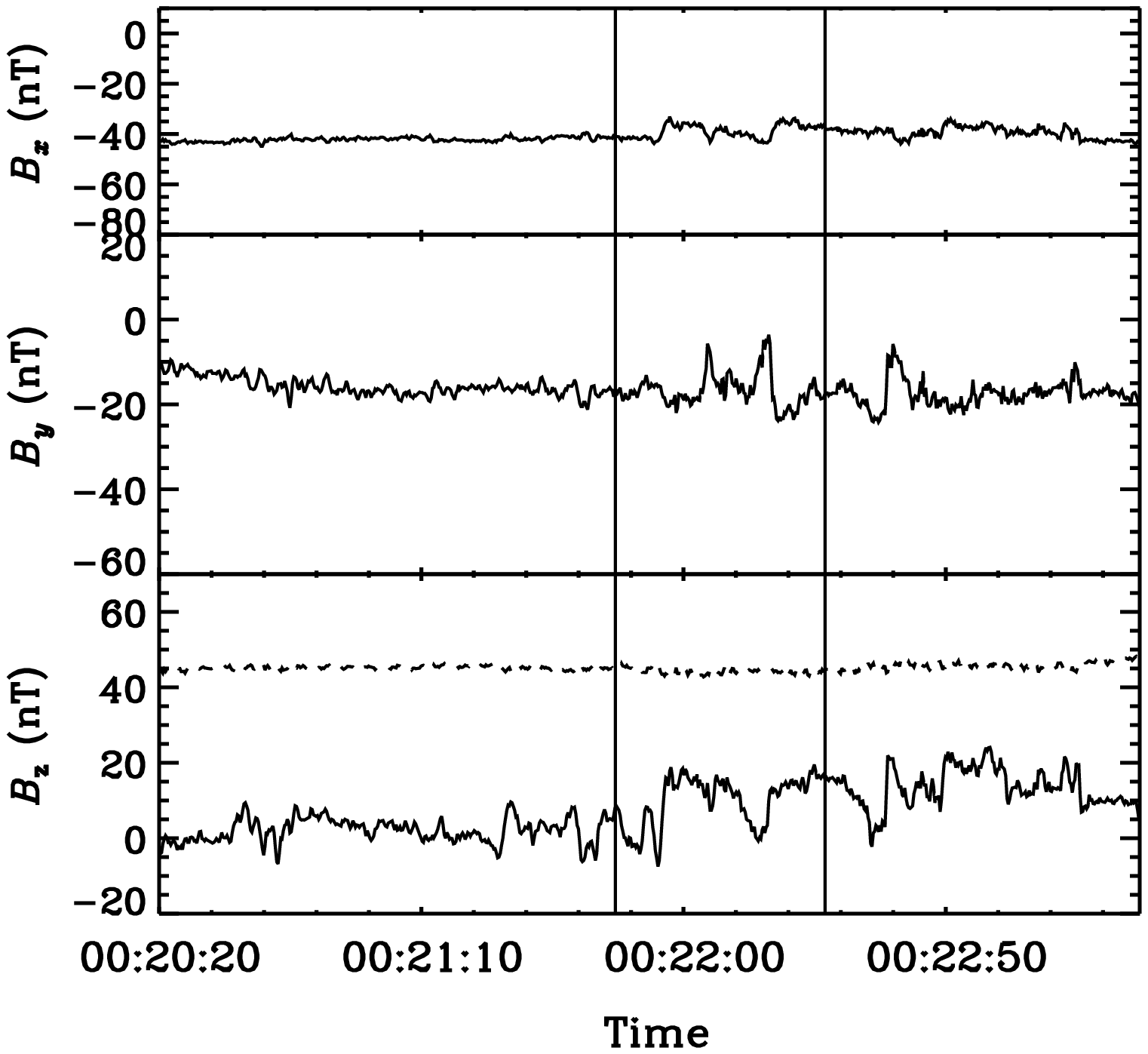}
\caption{Example of magnetic field measurements versus time in seconds on day
105 in 1976 at 0.29~AU, whereby the field components and magnitude are given
in solar-ecliptic coordinates, for which $X$ points toward the Sun, $Z$
towards the southern ecliptic pole, and $Y$ completes the left-handed
coordinate system. The magnitude of the full vector magnetic field (dashed line).} \label{fig.1}
\end{figure}

Obviously, direct and more adequate wave observations are required in order
to place new empirical constraints on the models and to limit the theoretical
assumptions. In the present work benefiting from the high-resolution magnetic
field data and the detailed proton VDFs obtained simultaneously by Helios~2,
we will present such observations and establish a clear correlation
between the proton temperature anisotropy and the power of high-frequency
transverse waves.

\section{Observations}

The data set used in this study was provided by the Helios~2 probe during its
primary mission, including the first perihelion passage in 1976 and covering
distances from 0.3 to 0.9~AU from the Sun. Here we consider simultaneous
measurements of the high-resolution 3D magnetic field vector components and
the proton VDFs. These measurements have been made by two different
instruments on board Helios 2, the magnetometer and the plasma experiment.
The high-resolution magnetic field data were obtained with a sampling
frequency that ranged up to values equal or higher than 4~Hz, corresponding to a Nyqvist-frequency
ranging up to values equal or higher than 2~Hz. This means that at all distances from the Sun, the field
sampling rate is high enough to cover the resonances near the local proton
cyclotron frequency, $f_p= \Omega_p/(2\pi)$.

The cadence of the plasma experiment for obtaining a VDF is typically 40.5~s,
yielding about 90 VDFs per hour in the case of continuous measurements
\citep[for more details see, e.g.,][]{marsch1982}. This cadence is longer
than the periods of the waves considered here. The Helios plasma experiment
does not separate the energy distributions of the proton core (the major
species) from the alpha particles and the proton beam. Since the
mass-per-charge ratio, $A/Z$, of the alpha particles is 2, we expect that
their VDFs should be clearly displaced from the maximum of the measured
combined ion VDFs, and thus the core of a VDF represents the bulk protons.
Since we are mainly interested in the core temperature anisotropy, we will
only consider the innermost part of any VDF, where it is higher than 20
percent of its maximum value.

In Figure~\ref{fig.1} we give a typical sample of the three components of the
magnetic field vector and its magnitude, as obtained on day 105 in 1976
during a period that starts at time 00:22:20 and lasts for 3 minutes. At this
time, the spacecraft was at a distance of 0.29 AU from the Sun. The magnetic
field data (in nT) are given in solar-ecliptic coordinates and reveal
the appearance of transverse (presumably Alfv\'en-cyclotron) waves with
substantial amplitudes (note the constant magnitude field in Figure~\ref{fig.1}).

\begin{figure}
\centering
\noindent\includegraphics[width=40pc]{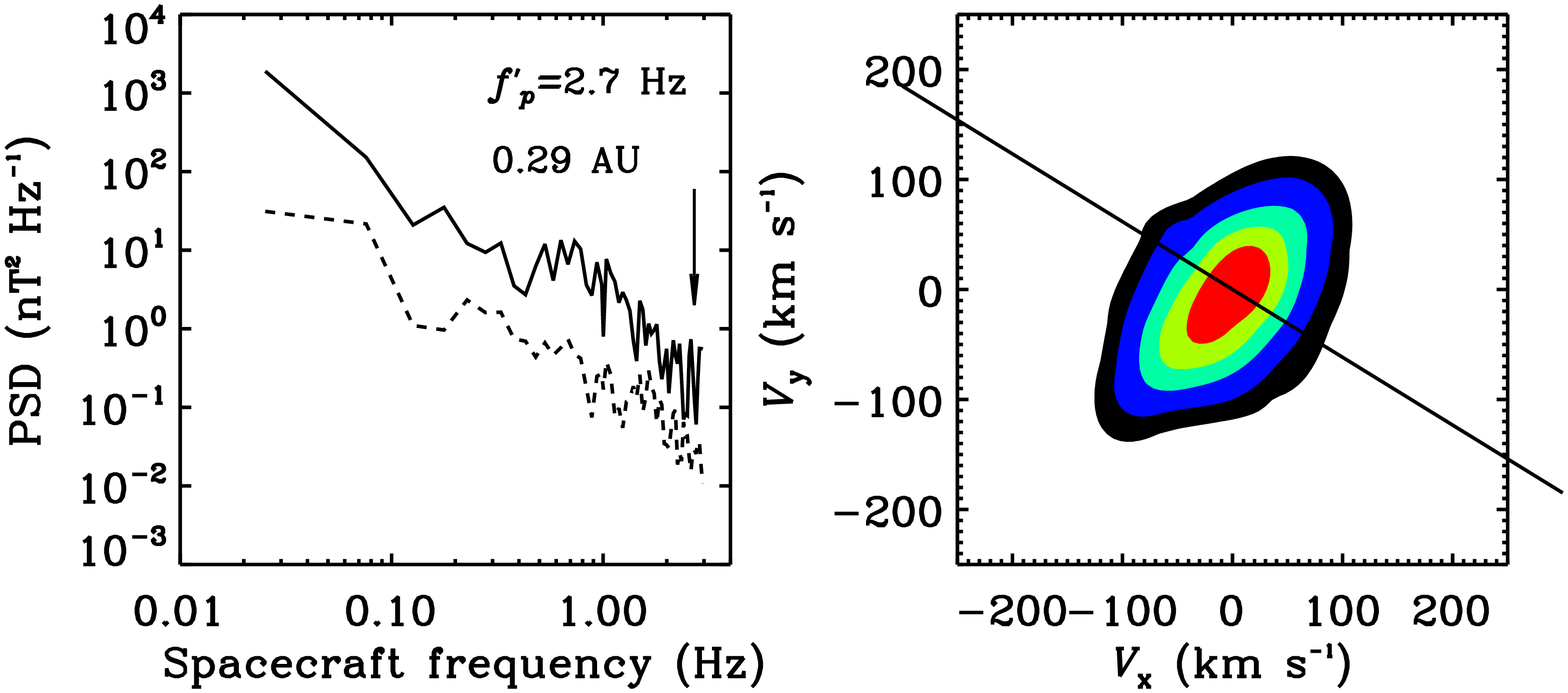}
\noindent\includegraphics[width=40pc]{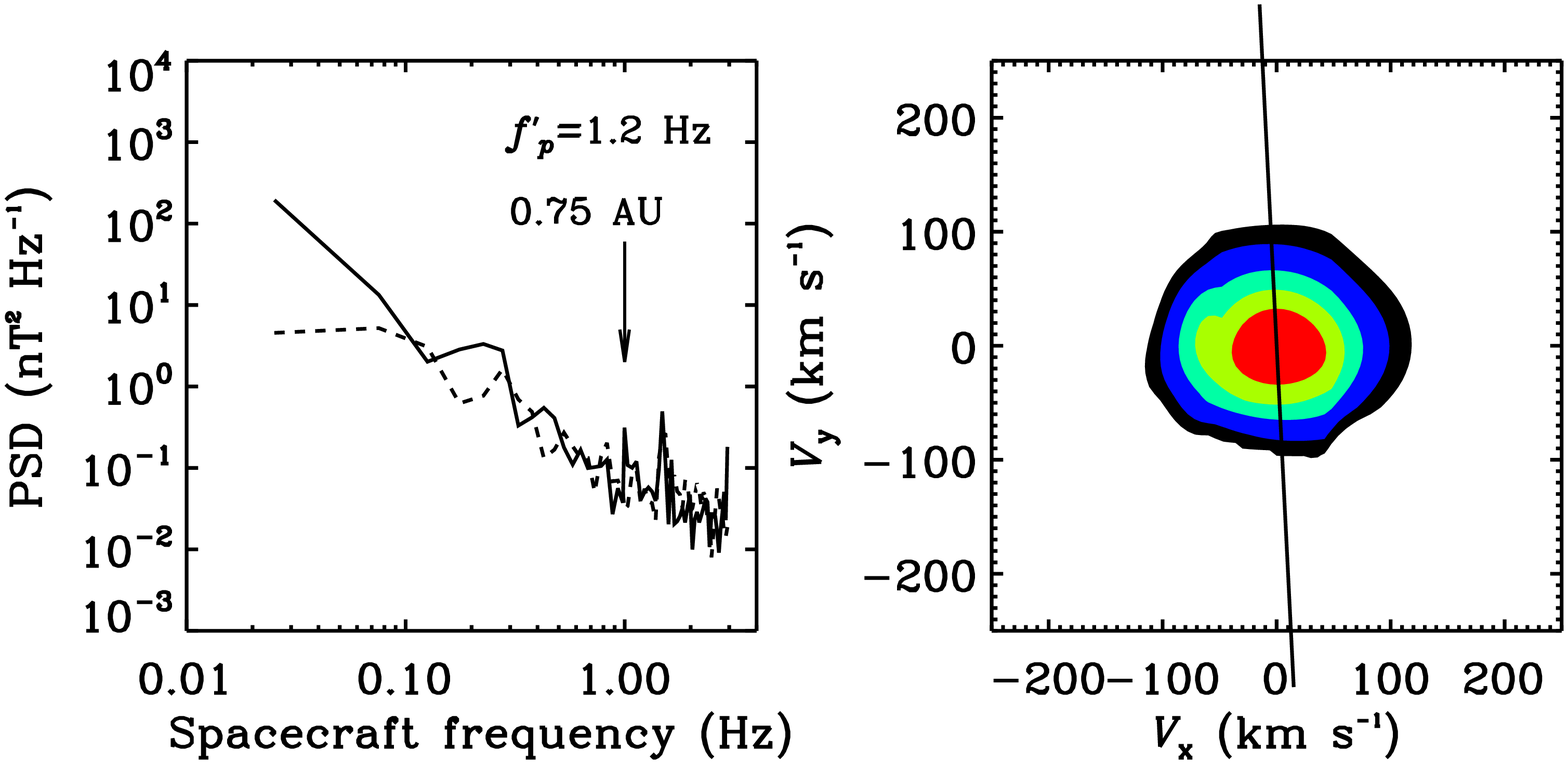}
\caption{Top panels: The
transversal (solid line) and longitidunal (dashed line) power spectrum (upper left panel)
of the waves during the interval time shown
between two solid lines in Fig.1, and the corresponding isodensity contours
of the proton VDF (upper right panel) measured on day 105 in 1976 at
00:22:07. Bottom panels: Another example showing the wave power spectrum
(lower left panel) during the time interval 21:36:50 and 21:37:32, and the
corresponding proton VDF (lower right panel) measured on day 67 in 1976 at
21:37:12. The Doppler-shifted proton cyclotron frequency $f'_p$ is marked by
an arrow in the left panels.} \label{fig.2}
\end{figure}

The left upper panel in Figure~\ref{fig.2} displays the power spectrum
density of the magnetic field turbulence measured during the time interval
bound by two vertical solid lines in Figure~\ref{fig.1}. Note that there are
peaks in the PSDs at a frequency of 1 Hz. These peaks are spurious and
consequences of the small remaining errors in the zero-offsets and the
spacecraft spin. However, in the case of the anisotropic VDF in the left top
panel of Figure~\ref{fig.2}, there is a power enhancement below the
Doppler-shifted proton cyclotron frequency, $f'_p=f_p(1+M_A)$, involving the
Alfv\'enic Mach number, $M_A=V/V_A$, of the solar wind with speed $V$. As we
can not measure the wave vector $\mathbf{k}$, we assumed here the maximal
possible Doppler shift for waves propagating outward from the Sun along the
local magnetic field direction. A wave energy density enhancement near the
cyclotron frequency has been found previously in other measurements, e.g. in
spectra obtained at 1~AU \citep{bale2005}.

In the right upper panel of Figure~\ref{fig.2}, we plot the corresponding
proton VDFs measured simultaneously during the same time interval when we
studied the magnetic field fluctuations. Another example of a power spectrum,
and the corresponding measured proton VDF, is plotted in the lower panel of
Figure~\ref{fig.2}. The data of this example were obtained on the day 67 in
1976 at the time 21:37:12, when the spacecraft was at a distance of about
0.75~AU from the Sun. It is worth noting that these projected 2D VDFs are
obtained from the 3D interpolation of the Helios plasma-experiment data. The
VDFs appear to be gyrotropic, with their axes of the symmetry representing
the direction of the background field (plotted as a solid straight line).

The wave power of the transversal component given in the upper panel
(solid line) is about one order of magnitude higher than that of the
longitudinal component (dashed line). Both components were computed
with respect to the mean field vector defined by the average over a time
interval of about 40~s. The transverse oscillations are composed essentially
of incompressible waves. Unfortunately, due to the lack of corresponding ion
bulk velocity fluctuations on that short time scale, it is not possible to
confirm if these magnetic fluctuations are of Alfv\'{e}nic nature. However,
most likely they are (note the constant magnitude field in Figure~\ref{fig.1}), as Alfv\'{e}n waves with periods larger than 40~s are
observed extensively in fast solar wind \citep{tu1995}.

Close inspection of the upper right panel of Figure~\ref{fig.2} reveals the
anisotropic shape of the core of the proton VDF. This characteristic form
indicates preferential ion heating in the perpendicular direction with
respect of the mean magnetic field. By comparing the two proton VDFs shown on
the right panels of Figure~\ref{fig.2}, we can see a distinct difference
between them. The one corresponding to higher wave power is more anisotropic
than the proton VDF corresponding to lower power. We conclude from
Figure~\ref{fig.2} that perpendicular proton heating is strongly correlated
with the power of transverse waves.

To provide more solid evidence for such a correlation between the proton
temperature anisotropy and the enhanced transverse wave amplitude, we made a
statistical analysis based on many measurements of both, high-resolution
magnetic field vectors and proton VDFs, which were obtained at different
locations from the sun. We selected 8 days of measurements in 1976, when the
spacecraft was at heliocentric distances ranging between 0.29 and 0.95~AU
from the Sun. Since our analysis only concerns the fast solar wind, we
preferred those measurement where the spacecraft was crossing fast streaming
plasma with a bulk speed higher than 600~km/s. Thus we selected several days,
with DOY numbers 36, 46, 51, 61, 67, 76, 95 and 105, on which the spacecraft
was, respectively, at the solar distances of 0.95, 0.90 ,0.87, 0.79, 0.75,
0.65, 0.41 and 0.29~AU.

In this study, we used the temperature ratio, $T_{\perp}/T_{\parallel}$, as
determined through computation of the parallel and perpendicular temperatures
derived from the velocity moments of the proton VDFs in 3D velocity space.
Also, we dealt with the average wave power spectrum obtained by
integration over the Doppler-shifted (indicated by a prime) frequency range,
(0.01--1)~$f^\prime$. Therefore, the mean magnetic field is obtained
by averaging the full magnetic field vector over a time interval of about
$ 10^{2}/f^\prime$.

Since the measurement cadence of the Helios plasma experiment is about 40~s,
we have more than 2000 measurements per day, and thus more than 16000
measurements for 8 days. Unfortunately, because the spacecraft did not
continuously encounter fast solar wind during the first perihelion passage,
there is a distance gap in the selected data, i.e., we do not have adequate
radial coverage between the distances 0.4 and 0.65~AU.

\begin{figure}
\includegraphics[width=40pc]{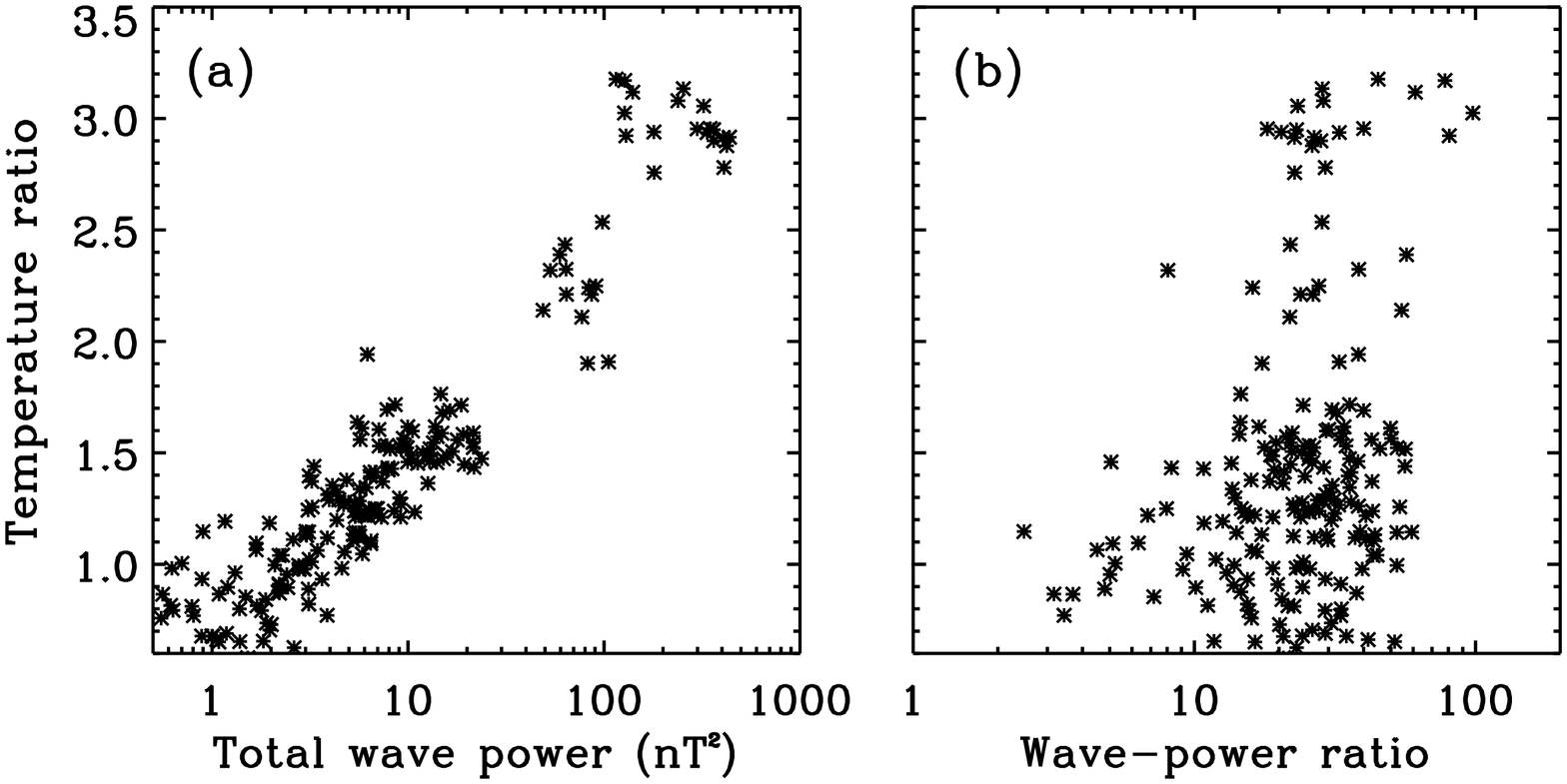}
\caption{(a): Mean temperature ratio, $T_{\perp}/T_{\parallel}$, versus mean
total wave-power (b): Mean temperature ratio, $T_{\perp}/T_{\parallel}$,
versus the ratio of the transversal to the longitudinal wave power. The
points correspond to the measurements made on days 36, 46, 51, 61, 67, 76, 95
and 105 in 1976.} \label{fig.3}
\end{figure}

\begin{figure}
\includegraphics[width=40pc]{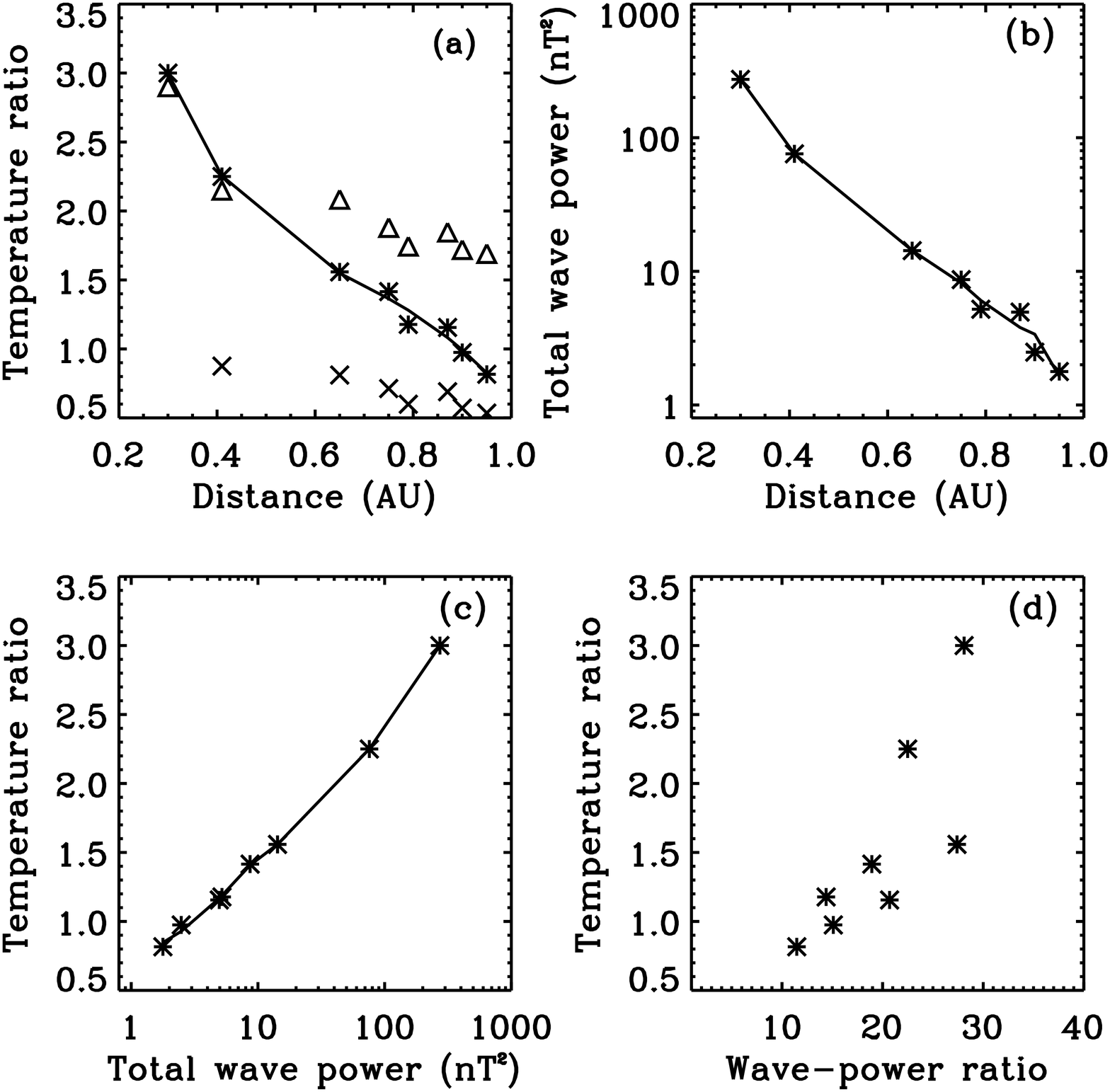}
\caption{(a): Mean temperature ratio, $T_{\perp}/T_{\parallel}$, plotted
versus distance from the Sun. A polynomial function was fitted to the data.
The thresholds of the temperature-ratio for the ion-cyclotron instability
(triangles), and for the firehose instability (crosses) are given for
different distances from the sun. (b): Mean total wave power versus distance
from the Sun, with a polynomial function fitted to the data. (c): Mean
temperature ratio versus mean wave power, with a five-order polynomial
function fitted to the data. (d): Mean temperature ratio plotted as a
function of the ratio of the transversal to longitudinal wave power.}
\label{fig.4}
\end{figure}

For the selected data set, we considered the mean values of the temperature
ratio and total wave power. These mean values are obtained through averaging
over each single hour of measurement time. Therefore, we have about 24 mean
values of each measured quantity for a day, and thus totally for all selected
data, we have about 192 average data points which are all plotted in
Figure~\ref {fig.3}. Overall, the temperature ratio spans values between
lower than 1 to a maximum of 3.5, and the total wave power ranges from low
values of about 0.5~nT$^{2}$ to high values of about 300~nT$^{2}$.

Notice that in Figure~\ref{fig.3}a an increase of the temperature ratio
corresponds to an enhancement of wave power. However, as it is known from
instability theory \citep{gary1993}, the temperature ratio tends to
be limited by a threshold value at which the plasma becomes unstable.
Furthermore, according to Figure~\ref{fig.3}b, the temperature ratio
increases with increasing ratio of transversal to longitudinal total wave
power. This latter ratio is mostly above 10 for temperature ratios above 1.5.

Considering the same data set, we now focus on the variation of the
temperature ratio and the total wave-power versus different distances from
the sun. The measured quantities given at each distance correspond to the
mean values for one day of measurements, and therefore, we have eight points
corresponding to the measurements made on days 36, 46, 51, 61, 67, 76, 95 and
105 in 1976 which are plotted as asterisks in Figure~\ref{fig.4}.

Both mean values, of the temperature ratio and total wave power, decrease
from their higher values close to the Sun to lower values at the distance of
0.9~AU far from the Sun. The temperature ratio declines from a value of 3 at
0.3~AU to about 0.6 at 0.95~AU. Correspondingly, the total wave power drops
from a value of about 250~nT$^{2}$ at 0.3~AU to about 2~nT$^{2}$ at 0.95~AU. The
transversal component of the mean wave power near the Sun is almost 30 times
higher than its longitudinal counterpart. Besides the total wave power, the
ratio of its transversal-to-longitudinal components is also a decisive factor
which seems to determine the observed variation of the temperature anisotropy
in fast solar wind.

From the linear theory of the plasma waves instability \citep[see
e.g.,][]{gary1993} it is possible to obtain threshold values of the
temperature anisotropy beyond which different wave modes will be excited,
e.g., the ion-cyclotron, mirror or firehose modes. Based on the parallel
plasma beta value (which ranges between about 0.1 and 0.6) we estimated
in Figure~\ref{fig.4} the typical threshold value of the temperature ratio
for the ion-cyclotron (diamonds) and firehose (crosses) instabilities. It
turns out that only at distances close to the Sun (0.3 and 0.4~AU), the value
of the temperature ratio is near the threshold of the ion-cyclotron
instability.

Apparently, due to the ion-cyclotron instability it seems not possible to
produce closer the Sun higher temperature ratios than the actually measured
ones. However, at distances greater than 0.5~AU from the Sun, the anisotropy
threshold causing the ion-cyclotron instability reduces to lower values,
mainly due to the radial increase of the plasma beta which for our
data clearly stays below unity. Since the wave-power decreases with
distance, the temperature ratios stay smaller than the threshold values for
the ion-cyclotron instability \citep{gary1993, tu2002}. But at distances
beyond 0.9~AU, the temperature ratio reaches values smaller than unity, and
then may hit the threshold for the firehose instability \citep{kasper2002}.

\section{Summary and discussion}

As shown above, a strong core temperature anisotropy,
$T_{\perp}/T_{\parallel}>1$, appears in the proton VDFs in fast solar wind
whenever the transverse magnetic field oscillations are enhanced. The
temperature anisotropy, as well as the power of these (very likely)
Alfv\'en-cyclotron waves ($\omega \leq \Omega_p$), both are at their highest
values at Helios perihelion distances near 0.3~AU. But at a farther distance
of about 0.9~AU the temperature ratio has decreased to values equal to or
below unity. Correspondingly, the wave power spectra decreased to much lower
values, about two order of magnitudes less than at 0.3~AU.

It has been shown that the high temperature anisotropy is correlated with the
enhanced transversal component of the waves in the frequency interval near
but below the local proton gyrofrequency. Alfv\'en-cyclotron waves appear to
have an important impact on proton heating and their resulting temperature
anisotropy. We were not able from the data we had to discriminate a
possible fast-magnetosonic component of these transverse waves. Their
longitudinal component is likely related to the magnetosonic slow mode wave
for our low-beta data.

Our findings support the notion that the ion heating process is connected
with large-scale plasma turbulence. The idea that high-frequency
ion-cyclotron waves are directly launched at the Sun might not be a valid
scenario. It seems more likely that these oscillations are generated
and reabsorbed continuously in the expanding solar wind through an ongoing
wave-particle cascade and regulated by temperature-anisotropy driven
instabilities as studied, e.g., by \cite{tu2002, kasper2002, bale2009} and/or
the parametric decay as suggested by the simulations of \cite{araneda2008,
valentini2009}.

%

\end{document}